\documentclass[12pt]{iopart} 

\usepackage[dvips]{graphicx}

\begin{document}

\title{An Axisymmetric Object-Based Search for a Flat Compact Dimension}

\author{Dylan Menzies and Grant J. Mathews}

\address{Department of Physics, University of Notre Dame}
\ead{dmenzies@nd.edu gmathews@nd.edu}

\begin{abstract}
A method is presented to search for a hypertorus symmetry axis by the alignment of distant objects. This offers greater sensitivity than previously proposed object-based methods that rely on accurate true distances. When applied to the catalog of objects with $z > 1$, we find no evidence for a compact dimension. We deduce a lower limit to the compact dimension size $D > 0.9$ of the distance to the cosmic horizon. This is consistent with independent constraints from the recent analysis of the WMAP microwave background data.
\end{abstract}

\maketitle

\section{Introduction}

Large compactified spatial dimensions often feature in modern theories such as String Theory, for example in scenarios where the dimensions are all initially small then some become large \cite{brandenberger89dimGrow,easther04dimGrow}. Whether the size of the compact space should be small enough for us to detect, however, is much less clear. Observations of the cosmic microwave background (CMB) and recent object surveys, have spurred interest in detecting evidence of a compact topology by looking for correlations in the positions of distant objects \cite{ lehoucq00crystalMethod,roukema96quintuplets,roukema02topologyConstraints}, or patterns in the CMB \cite{cornish98circles,levin01circleReview}. The strongest result appears to come from a recent analysis of the CMB \cite{cornish04circles}. From that work it has been deduced that for the hypertorus and many other topologies any compact dimension must be larger than $2 \cos(25^\circ)\eta_0 \approx 1.8 \, \eta_0$, where $\eta_0$ is the distance to the horizon from Earth (half the diameter of the horizon). There is, however, uncertainty in the validity of this method due to the thickness of the surface of last scattering from which the CMB originates. Some indications were previously found of spherical-dodecahedral topology, but these were ultimately not found to be statistically significant \cite{luminet04dodec,roukema04dodec,aurich04dodec,weeks04toriRatio}. The constraints from object based methods have been weaker \cite{lehoucq00crystalMethod, roukema96quintuplets, roukema02topologyConstraints}. Nevertheless, the motivation for a search for a hypertorus cosmology has been boosted by possible evidence of a suppression \cite{tegmark034cleanedWMAP} along a common axis in the quadrupole and octopole moments of the CMB. This suppression appears significant despite the increased cosmic variance at large scales \cite{oliveira03hypertorusTest}. While other explanations have been put forward \cite{piao03cmbSuppression,moroi04cmbSuppression,contaldi03cmbSuppression}, there remains the possibility that this suppression is due to a sub-horizon scale compact dimension, while taking into account the restrictions on dimension ratios introduced by \cite{weeks04toriRatio}. In particular we consider a compact dimension in flat space as produced by toroidal compactification, or as an approximation to slightly curved space \cite{veryflat}. For this case, we show that it is possible to construct a specialized object-based test with much greater sensitivity than previous object-based tests. This provides an independent approach to the CMB methods.

\section{Axisymmetric test}

\begin{figure}
\begin{center}
\includegraphics[width=5cm]{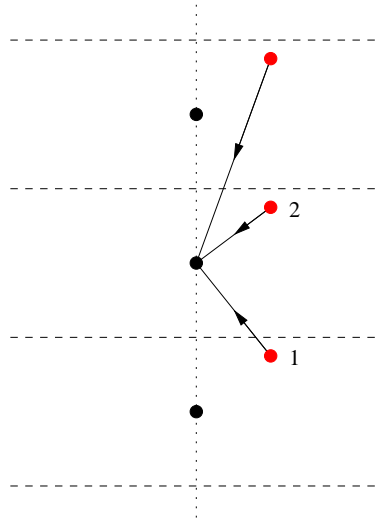}
\caption{\label{fig:geom1} Section through the comoving covering space showing light rays from the object to Earth}
\end{center}
\end{figure}

\begin{figure}
\begin{center}
\includegraphics[width=5cm]{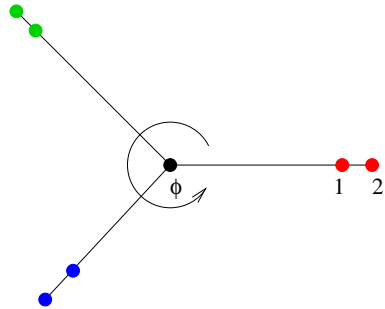}
\caption{\label{fig:geom2} Perspective view along the compact axis from Earth}
\end{center}
\end{figure}

The test is specific to cylindrical compactification. As discussed below, it can be modified for the other flat-space compactifications that arise from variations of the hypertorus, and even for curved space, but then the upper limit for the detected compact dimension is too small to be useful. Figure \ref{fig:geom1} shows a plane section through the comoving {\em covering space}, with the plane containing the compact dimension direction, represented by the dotted line. This is a way to visualize the geometry of the compact space, formed by imposing cyclic boundary conditions on a non-compact space, euclidean space in this case. The dashed lines are on 'copies' of a surface normal to the direction of compactification. Arrows represent different paths of light rays directed from a distant object to the Earth. The copies of an object all make the same angle $\phi$ around the compact dimension axis, shown as a dotted line. So, in a perspective projection along the axis from Earth, with front and back views superimposed, pairs of images (each corresponding to two copies of an object in the covering space) are each aligned with the center. Figure \ref{fig:geom2} illustrates this, with object numbering matching Figure \ref{fig:geom1}, and the orientation of $\phi$ indicated. Other object pairs are shown in different colors. This observation suggests that we look for an axis direction with an unusually high number of such alignments. The same idea appears to carry over to general topologies in constantly curved space, from observations using the {\em curvedSpaces} software\footnote{http://www.geometrygames.org/CurvedSpaces/}. However, the minimum separation of images that align in this way becomes a multiple of the shortest compact dimension, making it weaker than existing limits.

\begin{figure}
\begin{center}
\includegraphics[width=10cm]{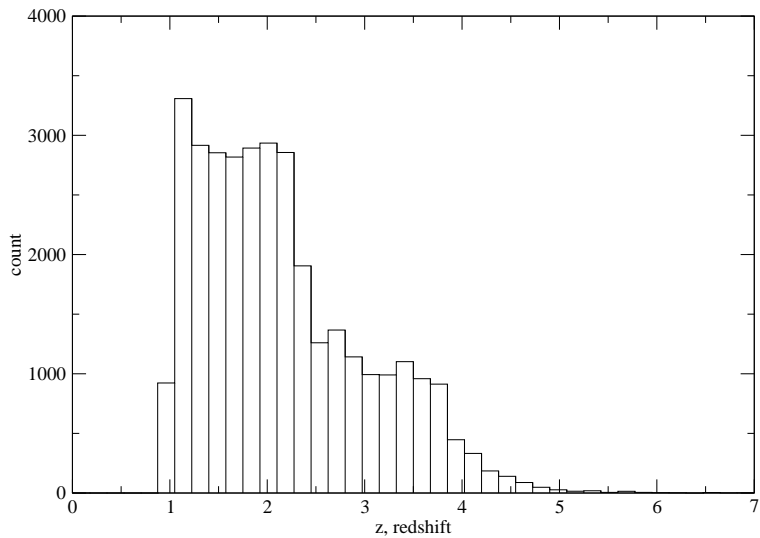}
\caption{\label{fig:NEDhistogram} Distribution of redshift among NED objects considered. }
\end{center}
\end{figure}

We shall now describe the test developed here in more detail. The objects used have $z > 1.0$, and are mostly quasars taken from the NASA/IPAC Extragalactic Database \footnote{http://nedwww.ipac.caltech.edu/} (NED). Nearer objects do not help in improving existing constraints on the hypertorus dimension. Figure \ref{fig:NEDhistogram} shows the range and distribution of redshift. A redshift of 5, where the data thins out, corresponds to a distance from earth of  $~0.5 \, \eta_0$, using the currently estimated $\Lambda$CDM cosmology. So we might to test a compact dimension up to a distance $(2)(0.5) \, \eta_0 = \eta_0$.

\begin{figure}
\begin{center}
\includegraphics[width=3cm]{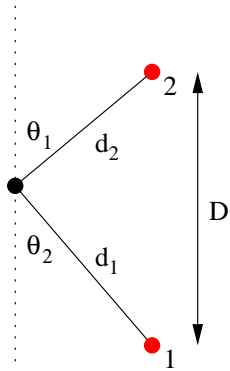}
\caption{\label{fig:geom3} Detail of section through covering space.}
\end{center}
\end{figure}

The Earth's peculiar velocity, its velocity relative to comoving space and the CMB, $\vec{\beta} \approx 0.00123$, causes a maximum angular error $\approx 0.1^\circ$ and maximum redshift error $\approx 0.01$ \cite{levin04aberratedCircles,lineweaver98CMBdipole}. This aberration is first removed from the survey data by applying a boost $-\vec{\beta}$ to incoming light rays. An aberrated line of sight vector $\hat{n}'$ is deaberrated \cite{deab} to a line of sight vector
\begin{equation}
\hat{n}= \frac{\hat{n}'+[(\gamma-1)\hat{\beta}.\hat{n}'-\gamma\beta]\hat{\beta}}
               {\gamma(1-\vec{\beta}.\hat{n}')} \quad \quad , \quad \gamma= (1-\beta^2)^{-1/2}
\quad .
\label{eq:n}
\end{equation}
Where $\hat{\beta}$ denotes $\vec{\beta}$ normalized.
Similarly, the observed redshift of an object $z'$ deaberrates to
\begin{equation}
z = \frac{(z'+1)}{ \gamma(1-\vec{\beta}.\hat{n}')} -1
\quad .
\end{equation}
The error remaining due to the motion of the Earth relative to the Sun, ~30 km/s, and the motion at the Earth's surface due to rotation, ~0.5 km/s, is from \cite{levin04aberratedCircles}, $\beta$ radians $\approx 0.006^\circ$. In order to detect a compact dimension of size up to the horizon, we look for image matches on opposite sides of the sky for each possible axis of the compact dimension. The alignment check is made more efficient by maintaining a list of objects ordered by angle, $\phi$, about the search axis. Alignments are then found by running through the list only once. After shifting the axis, the list is reordered slightly using a fast sorting algorithm. Due to the finite angular resolution of the search $\Delta\alpha$, each object is given an allowed range of $\phi$ values $[ \phi-\Delta\phi, \phi+\Delta\phi ]$ where $\Delta\phi = \Delta\alpha / \sin\theta$ and $\theta$ is the angle of the object away from the axis, as shown in Figure \ref{fig:geom3}. When the $\phi$ ranges of two objects on opposite sides of the axis overlap, a candidate match has been found.

\begin{figure}
\begin{center}
\includegraphics[width=10cm]{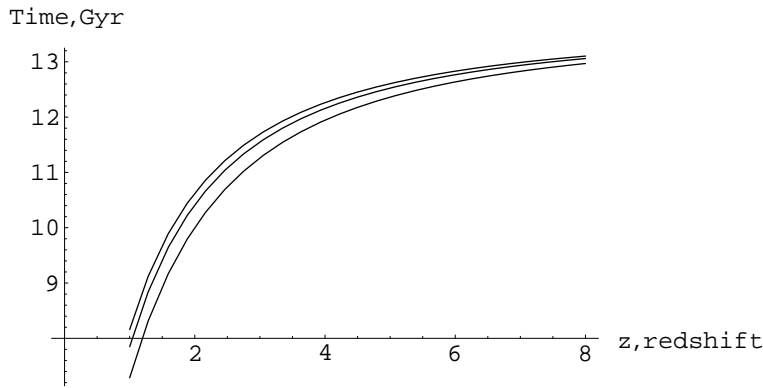}
\caption{\label{fig:tCurves} Look-back time functions for $\Omega_m = 0.17,0.3,0.42$(upper).}
\end{center}
\end{figure}

To reduce background noise from chance alignments, further tests are performed once an alignment has been found. Quasars are not expected to live longer than $10^9$ years \cite{roukema96quintuplets}, so images that differ in look-back time by more than this are discarded. The look-back time is determined from the redshift. We assume a flat $\Lambda$CDM model with $0.17 \le \Omega_m \le 0.42$. Variations in the look-back time caused by uncertainty in the cosmological model can be tolerated, since this has a minor effect on the rejection of pairs due to look-back time difference. Hence, a flat $\Omega_m = 0.3$ $\Lambda$CDM model is used. The similarity of the look-back time differences can be seen in Figure \ref{fig:tCurves}, which shows the curves for the extremes of $\Omega_m$ enclosing the $\Omega_m = 0.3$ curve. A function is used to accurately approximate the exact integral expression for look-back in a flat $\Lambda $CDM \cite{carrol92cosmoConstant} universe given by
\begin{equation}
t(z, \Omega_m, \Omega_{\Lambda}) = \frac{1}{H_0} \int_0^z dz \,(1+z)^{-1} (\Omega_m(1+z)^3 +\Omega_{\Lambda})^{-1/2}
\quad .
\end{equation}
Taking the age of the universe as 13.7 Gyr \cite{wmap}, the approximate look-back formula used with $\Omega_m = 0.3$ is
\begin{equation}
t(z) = 23.8(\frac{2}{3} (1-(1+z)^{-3/2}))-2.3 \quad  \textrm{Gyr .}
\end{equation} 

\begin{figure}
\begin{center}
\includegraphics[width=10cm]{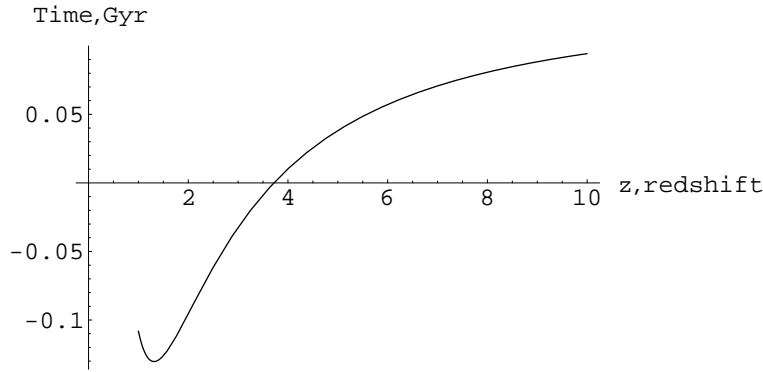}
\caption{\label{fig:tapproxError} Look-back time function approximation error.}
\end{center}
\end{figure}

This look-back time yields look-back time differences accurate to within 0.1 Gyr up to $z = 8$ for the given range of cosmologies, as shown in Figure \ref{fig:tapproxError}. The look-back filter also has the very positive effect of greatly reducing the angular error caused by the peculiar velocity of the object to a value which is well below other errors. This is because matching images at the same distance will originate at the same time, and hence from the same place in the compact space, regardless of the peculiar velocity. Finally, we notice that the perpendicular (covariant) distance from each image to the axis is the same, giving the constraint $\sin(\theta_1) d(z_1) = \sin(\theta_2) d(z_2)$. Allowing for angular and redshift errors, this translates into requiring an overlap of the two intervals,
\begin{displaymath}
[\sin(\theta_1-\Delta\alpha) d(z_1-\Delta z), \sin(\theta_1+\Delta\alpha) d(z_1+\Delta z)],
\end{displaymath}
\begin{displaymath}
[\sin(\theta_2-\Delta\alpha) d(z_2-\Delta z), \sin(\theta_2+\Delta\alpha) d(z_2+\Delta z)].
\end{displaymath}

We choose $\Delta\alpha = 0.01^{\circ}$ consistent with the residual errors from aberration and peculiar velocity, and $\Delta z = 0.01$ consistent with the observational error of many spectrographic redshift values. The test is sensitive to the cosmology determining the covariant distance function $d(z)$. Therefore it is repeated for a range of $\Lambda$CDM models about $\Omega_m = 0.3$. $d(z)$ is calculated using an interpolated lookup table for each cosmology. The tables are calculated using the following formula for distance \cite{carrol92cosmoConstant}, 
\begin{equation}
d(z) = \int_0^z dz ((1+z)^2 (1+\Omega_m z) - \Omega_\Lambda z(2+z))^{-1/2}) \quad ,
\end{equation}
normalized to 1 at the horizon. 

We also found that one final filter is necessary to exclude false peaks which result from tight clusters of correlated objects. These exist due to duplication of objects in the database and the high concentrations of objects found in pencil-beam surveys. These are excluded by organizing pairs into groups that are separated by a minimum angle of $0.5^\circ$. The angle was chosen because it is sufficient to remove the clustering effect, while giving a very small chance of reducing the numbers of genuine matching pairs.

Once a pair of images has passed all tests, the compact dimension size is estimated as $D = d_1 \cos \theta_1 + d_2 \cos \theta_2$, (cf. Figure \ref{fig:geom3}). The alignment is used to increment the bin corresponding to $D$ in a histogram corresponding to $\Omega_m$. The image references are also added to the bin, for possible later query. Once the search over $\phi$ is complete the histograms are analyzed to see if any bins contain more alignments than the existing record values. If so, the record values are updated, including image references. The search then proceeds to the next candidate axis. The HEALPix \footnote{http://www.eso.org/science/healpix/} function \verb#pix2ang_ring()#  \cite{healpix} is used to conveniently generate a uniformly spaced set of directions covering half the sky, and therefore all axes. The ring ordering ensures that all the angle increments are small, minimizing reorderings. The HEALPix resolution is chosen to be $N_{side} = 5000$ giving a pixel spacing $\Delta\alpha \approx 0.01^{\circ}$ consistent with the errors from the aberration and peculiar velocity.

\section{Simulated alignments}

\begin{figure}
\begin{center}
\includegraphics[width=10cm,trim=0 0 0 -3cm]{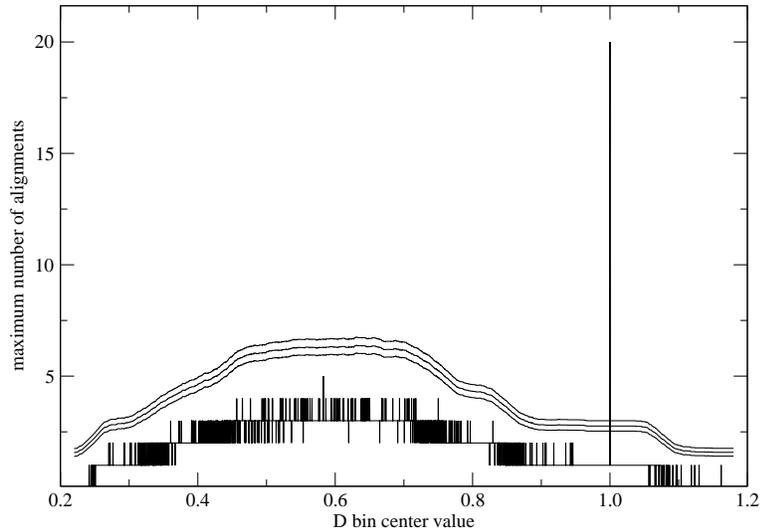}
\caption{\label{fig:simtest} Test of 20 simulated pairs with $D$ = 1.0, plus observed data.}
\end{center}
\end{figure}

As a test of the method and to calibrate the bin sizes we added simulated aligned objects to the database. The $D$ bin and $\Omega_m$ spacing are made just large enough so that a simulated set of aligned images, of look-back difference less than 1 Gyr, with maximum expected noise added and maximum misalignment between the $\Omega_m$ values, can be fully detected in one bin. The 26 cosmologies were tested were with $0.17 \leq \Omega_m \leq 0.42$ in steps of 0.01. The distance $D$, in units of distance to the horizon, was binned from 0.2 to 1.2 in steps of 0.002. The bins were overlapping to guarantee maximum detection rate. Figure \ref{fig:simtest} shows the output of record alignments with 20 simulated object pairs added to the objects data, and $D = 1.0$, for the closest cosmology tested. The additional lines are significance contours described in the next section.

\section{Results}

\begin{figure}
\begin{center}
\includegraphics[width=10cm,trim=0 0 0 -3cm] {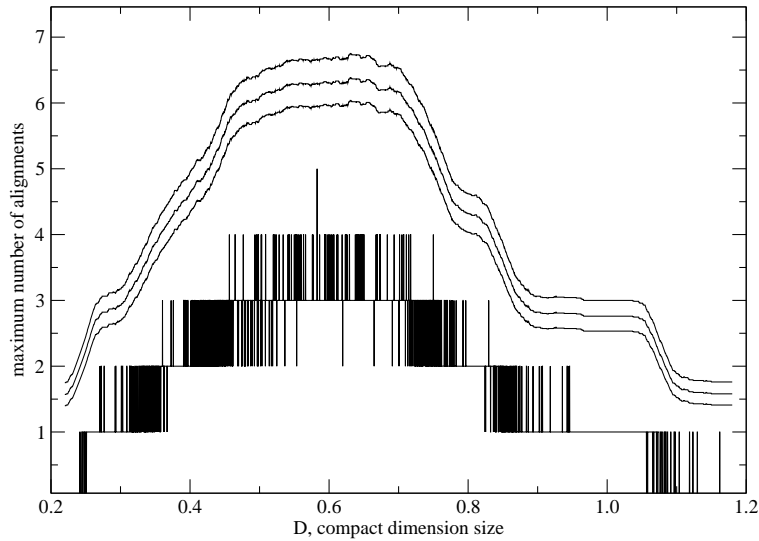}
\caption{\label{fig:results} Results of search compact dimension search for $\Omega_m = 0.3$ }
\end{center}
\end{figure}

The test was applied to a catalog of 40000 objects with $z>1.0$ obtained from the NED database. The search space was divided evenly between 20 1.4 GHz Linux workstations, and ran for 5 days. There was no clear evidence of a peak. To illustrate this, Figure \ref{fig:results} shows the maximum number of alignments found in each $D$ bin for $\Omega_m = 0.3$, once all the searches are combined. The contours are for $\sigma,2\sigma,3\sigma$ significance as described below. The cost of searching prohibits a detailed characterization of the noise statistics by repeated searches over decorrelated data sets. We expect only one anomalous peak, or possibly a few if we consider a hypertorus with more than one dimension inside the horizon, but all the plots show very similar local noise distributions. To quantify this more precisely we need to estimate the noise statistics from the result set and calculate how big a peak must be to be significant. We do this by looking at the fluctuation in alignments in each local region of $D$. The task is complicated by the number of alignments being a discrete variable. However, looking over the whole range of cosmologies we find that the local standard deviation in the noise, $\sigma_L$, varies from $\approx 0.7$ in the central regions to $\approx 0.3$ at the sides. The distribution for the maximum of a set of Poisson variables, which approximates the record alignments, falls off faster than a normal distribution. Therefore, we can make a conservative estimate of the {\em error contour} above which peaks are significant by using a normal model for the local noise distribution with $\sigma_L$. Let the contour be at $n \sigma_L$ above the local average of the noise. Then the probability of finding any peaks above the error contours for any of the cosmologies is $1-(1-Err_{RT}(n))^N \approx N Err_{RT}(n)$, where $N =$(num. cosmologies)(num. D-bins)$ = (26)(500) = 13000$ is the total number of bins in all cosmologies, and $Err_{RT}()$ is the right tail normal error function. To make this probability correspond to an overall $3\sigma$ significance we set it to 0.003, obtaining $n \approx 5.0$ and $n \sigma_L \approx 3.5$ in the central region. The contours corresponding to $\sigma,2\sigma,3\sigma$ significance are shown in Figure \ref{fig:simtest}. To summarize, the chance of finding any peaks above the $3 \sigma$ contour, for any of the cosmologies, is 0.003. We have given conditions for a statistically significant event. Next we need to show that the conditions will be met if a compact dimension actually exists.

\section{Sensitivity}

\begin{figure}
\begin{center}
\includegraphics[width=10cm, trim=0 0 0 -3cm]{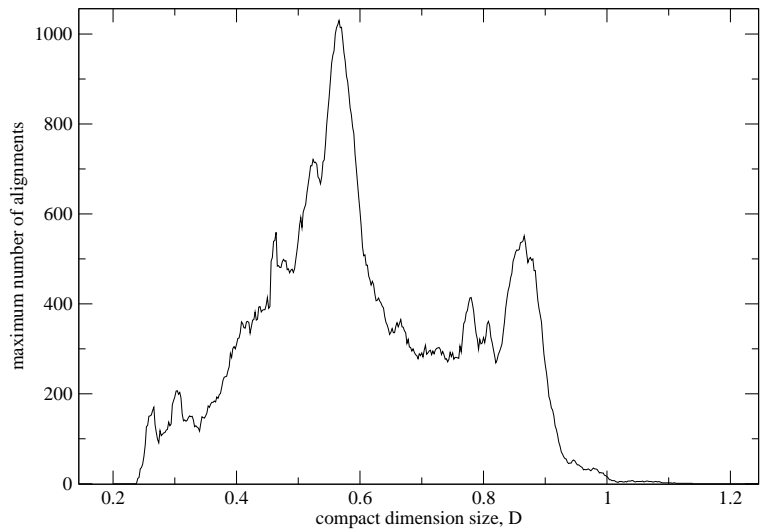}
\caption{\label{fig:completeAligns} Maximum alignments along one axis for the NED catalog made complete.}
\end{center}
\end{figure}

\begin{figure}
\begin{center}
\includegraphics[width=10cm, trim=0 0 0 -3cm]{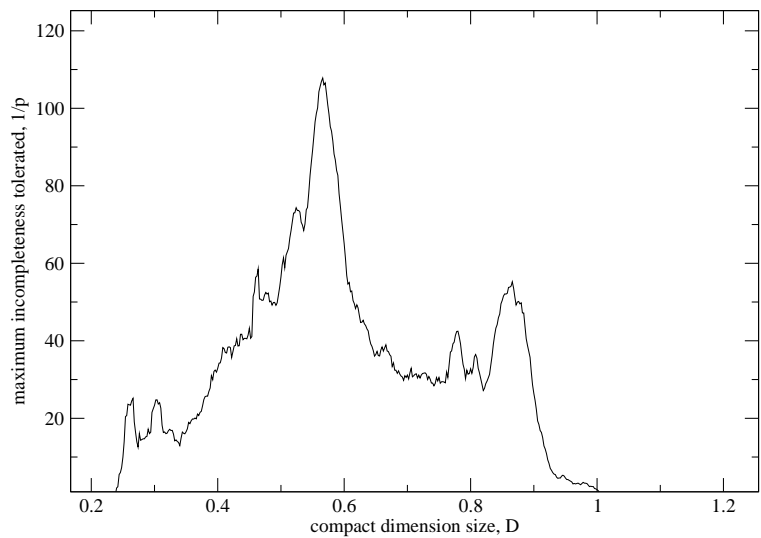}
\caption{\label{fig:sensitivity} Maximum incompleteness for which a significant detection can be made.}
\end{center}
\end{figure}

The question remains as to whether the test just described is sensitive enough to detect a compact dimension, and can we identify parameters in the catalog used that affect sensitivity. To answer this we first determined how many alignments would be generated along a given axis, for the full range of compact dimension sizes $D$, given observations on one half of the axis from the NED data, and assuming every observed image has a matching observable partner if they can satisfy the look-back time difference criteria. We define {\em completeness} as the fraction of those observed images that {\em could} have a matching partner satisfying look-back difference criteria, which actually do have an observed partner. Then the last assumption is equivalent to saying the completeness is 1, or we also say the data is {\em complete}. The result for a typical axis with $\Omega_m = 0.3$ is shown in Figure \ref{fig:completeAligns}. Figure \ref{fig:sensitivity} shows the ratio of the complete alignments to the $3\sigma$ error contour for $\Omega_m = 0.3$, which is the degree of {\em incompleteness}, the reciprocal of completeness, that can be tolerated while still making a positive detection. In other words from the initial calculation assuming completeness, we are calculating the lowest levels of completeness at which a detection can still be made for each $D$. For much of the $D$ range this is much higher than the incompleteness tolerated in \cite{lehoucq00crystalMethod}. In their work it is claimed that a toroidal topology can be detected with rejection of 90\% or completeness of $\approx 0.1$ in the current catalog. However, this rejection rate includes rejections due to look-back time difference, whereas in our treatment the completeness quoted is for the original data set. In practice look-back difference rejection would reduce the overall completeness well below their rejection threshold, so that even with a complete original set their test would not be sensitive enough.

The next question is can we estimate the incompleteness of our data? One way to approach this is to look at the density of objects in detailed pencil surveys as an indication of total number of objects we can detect, and compare this with the average across the sky. Doing this for a few pencil surveys gives a density ratio of $\approx 10^4$ implying that the data is too incomplete to detect a compact dimension. However this probably overestimates the incompleteness because the pencil surveys include objects of much lesser magnitude than the remaining surveys, and we expect matched objects to be of similar absolute magnitude, because of similar look-back times. So, it is possible that the survey is complete enough to make detection possible over a wide range of $D$: From Figure \ref{fig:sensitivity}, a completeness of $0.02$ would rule out a compact dimension up to $0.9 \, \eta_0$.

The dependence of sensitivity on completeness can be modeled as follows. Suppose a complete observable set contains $n$ matched pairs, and the fraction of images actually observed or completeness is $p$. Then $n p^2$ matched pairs will be observed on average. The number of pairs observed by chance can be checked empirically to be of the form $\nu p^2$, for a constant $\nu$. This can be explained as the maximum of a set poisson distributions, which is proportional to the mean of a single poisson distribution describing the number of alignments in a given direction. The mean is clearly proportional to $p^2$ since an alignment occurs when events on two linear poisson processes intersect. The upshot is, perhaps surprisingly, that the ratio of signal to noise $n / \nu$ is independent of $p$, the completeness, except that additionally we require $p$ sufficient to ensure that the signal is separated from the fluctuations in the noise. This means that we are immune to incompleteness. The calculation of incompleteness previously was used to establish the range of $D$ to which we are sensitive, and this range will remain valid for higher levels of completeness. It is therefore unnecessary to process more than a certain number of images, a random selection suffices. The only way to improve sensitivity is by choosing more selection criteria that can raise $n/\nu$. This will reduce the detected signal and noise, but may require more objects so that they can be separated.

\section{Summary}

The approach described is more sensitive than previous object-based tests applied to the hypertorus, because it does not rely on accurate true distance determination. Although there is uncertainty over the degree of completeness among the observations, the results presented here suggest that a hypertorus cosmology is ruled out for $D < 0.9 \, \eta_0$. Selection could be improved by matching magnitude and spectral profiles, using greater an enlarged catalog. Deeper observations going beyond the current $\approx 0.5 \, \eta_0$ limit will extend the range of $D$ tested. \\

Thank you to the reviewer for numerous suggestions that improved the readability of the manuscript. Work at the University of Notre Dame supported by the U.S. Department of Energy under research grant DE-FG02-95-ER40934, and a University of Notre Dame Center for Applied Mathematics (CAM) summer fellowship. This research has made use of the NASA/IPAC Extragalactic Database (NED) which is operated by the Jet Propulsion Laboratory, California Institute of Technology, under contract with the National Aeronautics and Space Administration. Some of the results in this paper have been derived using the HEALPix \cite{healpix} package.

\end{document}